\documentclass{article}

\usepackage{arxiv}

\usepackage[utf8]{inputenc} 
\usepackage[T1]{fontenc}    
\usepackage{hyperref}       
\usepackage{url}            
\usepackage{booktabs}       
\usepackage{amsfonts}       
\usepackage{nicefrac}       
\usepackage{microtype}      
\usepackage{amsmath}

\usepackage{xcolor}
\usepackage{soul}
\definecolor{Green}{HTML}{07540e}

\usepackage{graphicx}

\title{Urban Scaling of COVID-19 epidemics}

\author{
  Ben-Hur Francisco Cardoso \\
   Instituto de Física - UFRGS\\
  \texttt{ben-hur.cardoso@ufrgs.br} \\
   \And
 Sebastián Gonçalves\\ 
   Instituto de Física - UFRGS\\
  \texttt{sebastiangoncalves@gmail.com} \\
}

\begin{document}
\maketitle

\begin{abstract}
Susceptible-Invective-Recovered (SIR) mathematical models are in high demand due to the COVID-19 pandemic. They are used in their standard formulation, or through the many variants, trying to fit and hopefully predict the number of new cases for the next days or weeks, in any place, city, or country. Such is key knowledge for the authorities to prepare for the health systems demand or to apply restrictions to slow down the infectives curve.
Even when the model can be easily solved ---by the use of specialized software or by programming the numerical solution of the differential equations that represent the model---, the prediction is a non-easy task,  because the behavioral change of people is reflected in a  continuous change of the parameters.
A relevant question is what we can use of one city to another; if what happened in Madrid could have been applied to New York and then, if what we have learned from this city would be of use for São Paulo.
With this idea in mind, we present an analysis of a spreading-rate related measure of COVID-19 as a function of population density and population size for all US counties, as long as for Brazilian cities and German  cities. Contrary to what is the common hypothesis in epidemics modeling, we observe a higher {\em per-capita} contact rate for higher city's population density and population size. Also, we find that the population size has a more explanatory effect than the population density. A contact rate scaling theory is proposed to explain the results.
\end{abstract}

\keywords{COVID-19, Epidemics, Urban scaling, SIR model}

\section{Introduction}
The epidemic of COVID-19 that started in the Chinese city of Wuhan in December of 2019, was declared a pandemic on March 11th, 2020 by the World Health Organization (WHO). Presently, the epicenter is in the US, while cases still are growing in many European countries. Soon, it will shift its center of gravity to Russia or Brazil, where the epidemic has the potential to hit even worse than in the US.

Many groups still struggle to get precise median and long term predictions of the number of expected cases, especially hospitalized or ICU ones. Such is because the epidemic parameters are continuously changing as the population changes its behavior, with or without government interventions~\cite{Weber2020}.
However, some general properties can already be identified across aggregated data, specifically related to demographic characteristics.

To give ground to our proposed analysis, which is to compare empirical municipal-level data in some countries, we use a version of the SIR model, which we call the SIRD model, because of the fourth $D$ compartment. 
In this simple model, each municipal-level region (city or county) with population size $N$ and land area $A$ is composed of the following epidemiological compartments: susceptible ($S$), infected ($I$), recovered ($R$), and dead ($D$). Considering only within-county transmission, the dynamics of these compartments is driven by the following system of differential equations~\cite{keeling2011modeling}:

\begin{align}\label{eq:sir}
\nonumber\frac{dS}{dt} &= -\frac{\beta}{N} I S\\
\nonumber\frac{dI}{dt} &=  \frac{\beta}{N} I S - \gamma I\\
\nonumber\frac{dR}{dt} &= (1 - \phi)\gamma I\\
\frac{dD}{dt} &= \phi \gamma I,
\end{align}

where $\beta$ is the transmission rate, $\gamma$ is the removal rate, and $\phi$ is the case fatality rate. 
The $N$ factor in the denominator makes $\beta$  a disease only parameter, supposedly independent of the size or other characteristic of the population.
Indeed, in the book of Keeling and Rohani~\cite{keeling2011modeling} this formulation is referred to as frequency-dependent (or
mass action) transmission. Called it proportionate mixing by Anderson and May~\cite{anderson-may1991}, it assumes that the number of contacts is independent of the population size, resulting in similar patterns of transmission, whether it is a town or a large city.
However, in the unprecedented evidence that we are collecting from the ongoing COVID-19 pandemic, that common intuition seems not to be generally valid. 
On the opposite side, there is the pseudo-mass action formulation~\cite{keeling2011modeling}, in which the infection rate is directly proportional to the population size ---which is not usually applied to human infectious diseases. Our analysis shows that none of these extreme formulations can satisfactorily explain the available COVID-19 data. The best fit corresponds to a formulation that is somehow in between those ones, and which can be explained in terms of a contact rate scaling theory.

\section{Population size and Population density}
The time evolution of these compartments is governed by the three parameters, $\phi$, $\gamma$, and $\beta$.
The last one can be factorized as $\beta  = pC$,  where $p$ is the probability of transmission and $C$ is the {\em per capita} contact rate~\cite{hu2013scaling}. The probability of infection $p$~\footnote{It can be drastically attenuated or even suppressed  by the use of masks, for example.} is a characteristic of the disease, most likely universal, and a key to epidemics because if it is too low, we would probably not have an outbreak.

$C$, on the other side, condenses all the human factors that give rise to different epidemic patterns in different places, countries, or cultures. It is the only parameter that non-pharmaceutical interventions, like activity restrictions or lock-downs, can modify. Yet, we will restrict ourselves here to its urban dependency.
There are two main competing hypothesis that try to explain how $C$ varies with $N$ and $A$: the population size driven contact rate, where $C = C(N)$; and the population density driven contact rate, stating that $C = C(\rho)$, where $\rho = N / A$. While the first approach assumes that the social mobility network grows in larger areas, allowing more distant people to interact~\cite{bettencourt2013origins}, the second one assumes that the length traveled by the individuals is invariant of the city's size~\cite{pentland2020}.

\begin{figure}[!htb]
	\centering	
	\includegraphics[height=7cm]{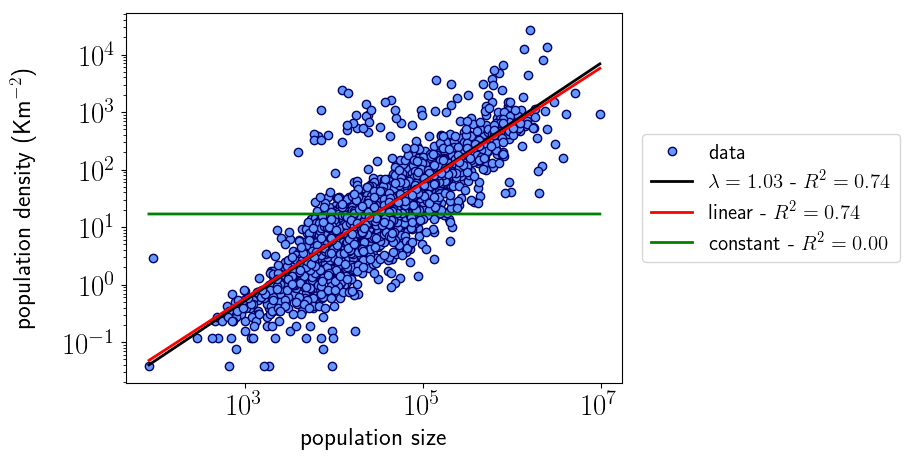}
	\caption{Population density ($\rho$) {\em vs} population size ($N$) for US counties, represented by blue dots (data). The black line is a power law fitting $\rho \propto N^{\lambda}$, while the red line is a linear fitting $\rho = kN$. Both have equal correlation coefficient, $R=0.86$. The green line represents  the average density. See Section~\ref{data} for the data source.}
	\label{fig:pop_den_US}
\end{figure}

Intriguingly, based on data of disease transmission in the United States, both approaches appear to be valid~\cite{hu2013scaling, pentland2020, bjornstad2002dynamics}.
The reason for this is the quasi-linear correlation between density and size population of the US's counties, as shown in Fig.~\ref{fig:pop_den_US}.
Indeed, we have found that the best fit is $\rho \propto N^{\lambda}$, with $\lambda \approx 1.03$. Assuming a linear relation instead, $\rho = kN$ gives an equally valid fit, where $k = A^{-1} = 0.00059km^{-2}$. An almost constant density across counties, or no correlation between  $\rho$ and $N$ cannot explain the data.
Note that from the value of the constant $k$ we can obtain a typical county diameter in the United States of $46.5km$.

\begin{figure}[!htb]
	\centering	
	\includegraphics[width=0.48\textwidth]{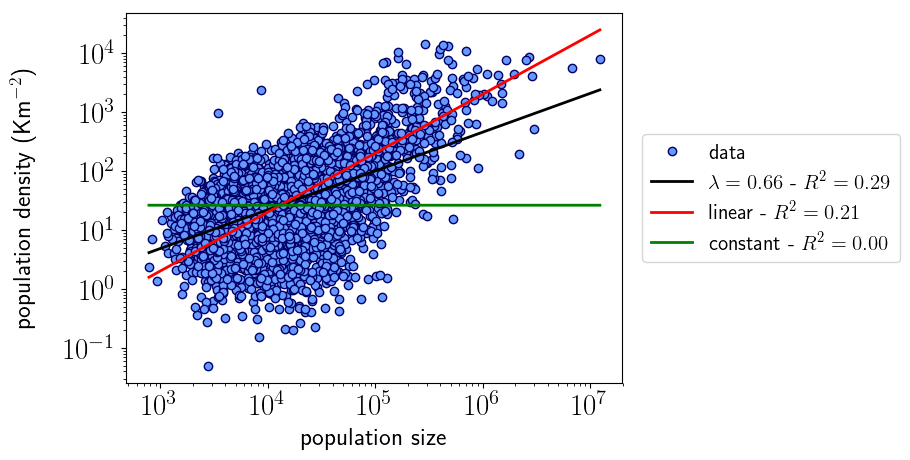}
	\includegraphics[width=0.48\textwidth]{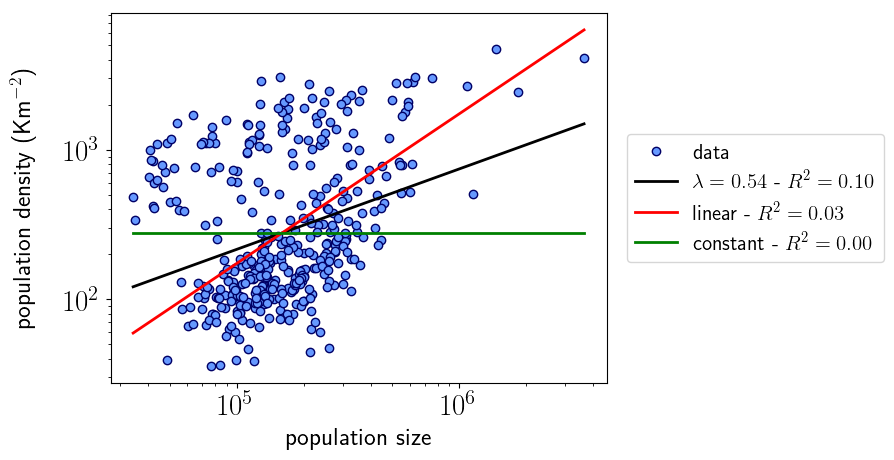}
	\caption{Population density ($\rho$) {\em vs} population size ($N$) at the municipal level for Brazil (left) and Germany (right), represented by blue dots (data). Red lines are linear fitting $\rho = kN$, with $k=0.002$ for Brazil and $k = 0.0017$ for Germany. The green lines corresponds to the average density of each country.
	See Section~\ref{data} for the data source.}
	\label{fig:pop_den_BR_GE}
\end{figure}

In addition to the US, we study the COVID-19 transmission in Brazil's and Germany's cities. In these two countries, the city's population size does not correlate well with their population density (see Fig.~\ref{fig:pop_den_BR_GE}). The linear fitting, $\rho = kN$, is weak for the Brazilian cities and almost nonexistent for Germany.
The constant case also cannot explain the data. Since there is no correlation between $\rho$ and $N$, we can use these two countries to check the validity of the population size-driven or the population density-driven approaches. The results can be useful during the present COVID-19 pandemic and for futures ones.

\section{Contact rate scaling theory} 
Let assume individuals distributed uniformly in a two-dimensional space according to a density $\rho$. As introduced by Noulas {\em et al}~\cite{noulas2012tale}, we can expect that the individual $j$ interacts with the individual $i$ with probability
$$P_{ji} = \bigg(\frac{1}{\text{rank}_i(j)}\bigg)^{1-\alpha},$$
where $\text{rank}_i(j)$ is the number of neighbors closer to $i$ than $j$ and $0 \leq \alpha \leq 1$ is a scaling factor. Assuming that the distance between these two individuals is $r$, we have that
$$\text{rank}_i(j) = \rho \pi r^2 \Rightarrow P_{ji} = P(r)= \bigg(\frac{1}{\rho \pi r^2}\bigg)^{1-\alpha}.$$

First, since $0\leq P \leq 1$, we must impose a bottom cutoff radius $r_0$ such that $$P(r \leq r_0) = 1 \Rightarrow r_0 \equiv \frac{1}{\sqrt{\pi \rho}}.$$
Secondly, it is natural to assume an upper cutoff radius $r_1$ for long distances such that $P(r > r_1) = 0$. So, the {\em per capita} contact rate is given by
\begin{equation}
\label{eq:C}
C_{\alpha} = \frac{1}{2}\int dr \> (2\pi r\rho) P(r) = \frac{1}{2}\bigg[\pi \rho r_0^2 + \int_{r_0}^{r_1} dr \> (2\pi r\rho) P(r)\bigg] \Rightarrow C_{\alpha} = \frac{(A_1 \rho)^{\alpha} +\alpha - 1}{2\alpha},
\end{equation}
where $A_1 \equiv \pi r^2_1$ is the coverage area of individual mobility and the $1/2$ factor eliminate the double counting. This result generalizes the $\alpha = 0$ case deduced by Krumme {\em et al}~\cite{pentland2020}, where 
$$C_0 = \lim_{\alpha \rightarrow 0} \bigg[\frac{(A_1 \rho)^{\alpha} +\alpha - 1}{2\alpha}\bigg] = \frac{1 + \ln(A_1 \rho)}{2}.$$

According to the population size-driven approach, the mobility increases with $A$; $A_1 = \kappa A$, then from Eq.~\ref{eq:C} we have 
$$C_{\alpha} = C_{\alpha}(N) = \frac{(\kappa N)^{\alpha} +\alpha - 1}{2\alpha}.$$ 
On the other hand, the population density driven approach states that $A_1$ is invariant, thus from the same Eq.~\ref{eq:C} we got
$$C_{\alpha} = C_{\alpha}(\rho) = \frac{(A_1 \rho)^{\alpha} +\alpha - 1}{2\alpha}.$$ 

\section{Data}\label{data}
We use the municipal-level time series of confirmed cases and deaths for United Sates~\cite{CSSEGI}, Brazil~\cite{brasilio} and Germany~\cite{jgehrcke}. 

Also, we use the municipal-level population size and land area for United Sates~\cite{eua_pop,eua_area}, Brazil~\cite{brasil_pop, brasil_area} and Germany~\cite{destatis}.

\section{Methodology}
\begin{figure}[!htb]
	\centering	
	\includegraphics[height=7cm]{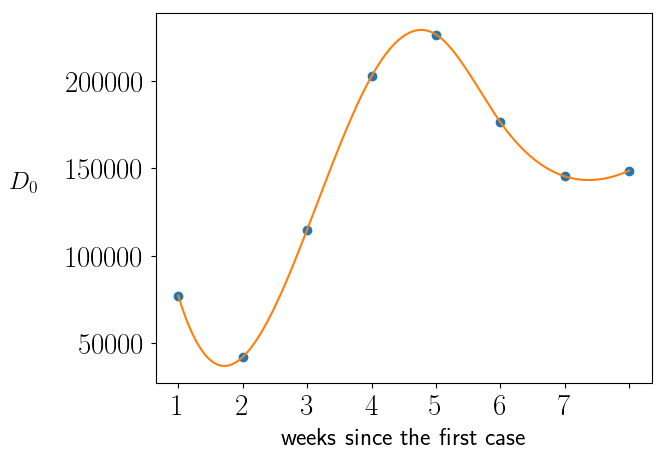}
	\caption{Weekly evolution of $D_0$ of the New York county of US. We can note the maximum value in the 5th week, at the beginning of the PAUSE order~\cite{pause-ny}.}
	\label{fig:time_d0}
\end{figure}

Due to social distancing measures, it is expected that the value of $\beta$ varies in time, but we can consider it as a constant for a sufficient short interval. So, let be $[t, t + \Delta t]$ such that $S(t+\Delta t) < S(t)$ and  $D(t+\Delta t) > D(t)$. Assuming that $\beta(t)$ is constant in this interval, we get form Eq.~\ref{eq:sir}:
$$ \frac{dS}{dD}  =  \frac{dS}{dt}  \bigg/\frac{dD}{dt} = -\bigg(\frac{\beta}{N\phi \gamma}\bigg)S\Rightarrow S(D) = Be^{-D/D_0},\>\>D_0 \equiv \frac{N\phi \gamma}{\beta},$$
where $B$ is the integration constant. Aiming to cancel some day of week seasonality bias, we choice $\Delta t = 1$ week. Now, noting the $S$ is the population size minus the confirmed cases, we can construct a weekly time-series of $D_0$ (see Fig.~\ref{fig:time_d0} for an example) such that
$$\frac{S(t)}{S(t+ \Delta t)} = e^\frac{D(t+\Delta t) - D(t)}{D_0} \Rightarrow D_0(t) = \frac{D(t+\Delta t) - D(t)}{\ln S(t) - \ln S(t+\Delta t)}.
$$

Assuming that the social distancing measures are sufficiently distributed among cities and time, we can expect that $D_0$ varies in time around a value proportional to the theoretical one.

With the population density-driven contact rate hypothesis, we relate the population density of each city with its $\langle D_0 \rangle / A$ value, where $\langle D_0 \rangle$ is the time average of $D_0(t)$ for this location. So, in this framework, we have

\begin{equation}
\label{eq:model_den}
\frac{\langle D_0 \rangle}{A} \propto \bigg(\frac{\phi \gamma}{p}\bigg)\frac{\rho}{C(\rho)} \propto \frac{\alpha \rho}{(A_1 \rho)^{\alpha} +\alpha - 1}.
\end{equation}

Considering the population size-driven contact rate hypothesis, we plot the population size of each city and its $\langle D_0 \rangle$, where $\langle D_0 \rangle$ is the time average of $D_0(t)$ for this location. In this approach, we expect

\begin{equation}
\label{eq:model_pop}
\langle D_0 \rangle \propto \bigg(\frac{\phi \gamma}{p}\bigg)\frac{N}{C(N)} \propto \frac{\alpha N}{(\kappa N)^{\alpha} +\alpha - 1}.
\end{equation}

\section{Results}
\subsection{Population density driven contact rate}

If the mass action (constant contact rate) is valid, we expect that $\langle D_0 \rangle / A$ scales linearly with $\rho$. In the another extreme, if the pseudo-mass action ($C \propto N$) is valid, we expect that $\langle D_0 \rangle / A$ is a constant.

\begin{figure}[!htb]
	\centering	
	\includegraphics[height=7cm]{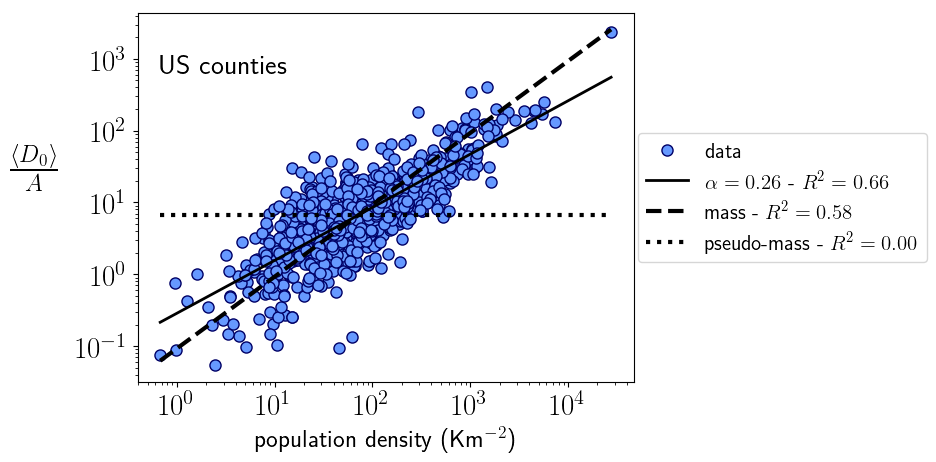}
	\caption{Comparison between $\langle D_0 \rangle / A$ and the population density for different counties of United States.}
	\label{fig:US_rho}
\end{figure}

\begin{figure}[!htb]
	\centering	
	\includegraphics[height=7cm]{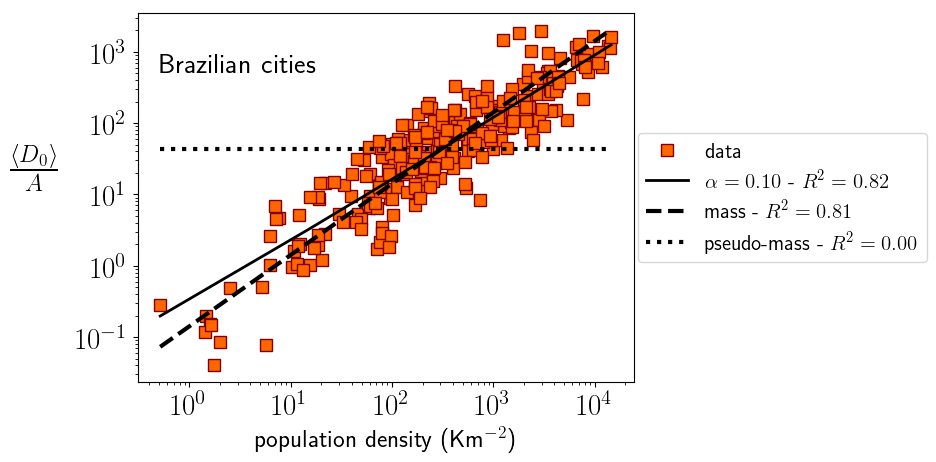}
	\caption{Comparison between $\langle D_0 \rangle / A$ and the population density for different cities of Brazil.}
	\label{fig:BR_rho}
\end{figure}

\begin{figure}[!htb]
	\centering	
	\includegraphics[height=7cm]{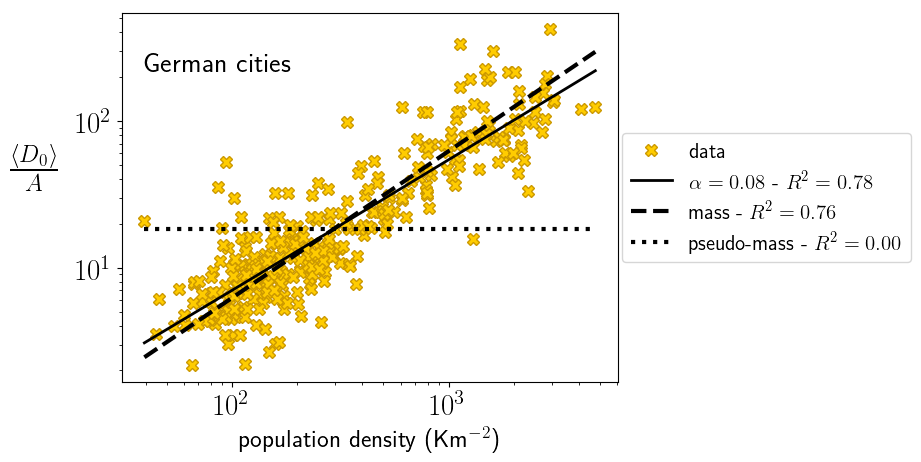}
	\caption{Comparison between $\langle D_0 \rangle / A$ and the population density for different cities of Germany.}
	\label{fig:GE_rho}
\end{figure}

\begin{figure}[!htb]
	\centering	
	\includegraphics[height=7cm]{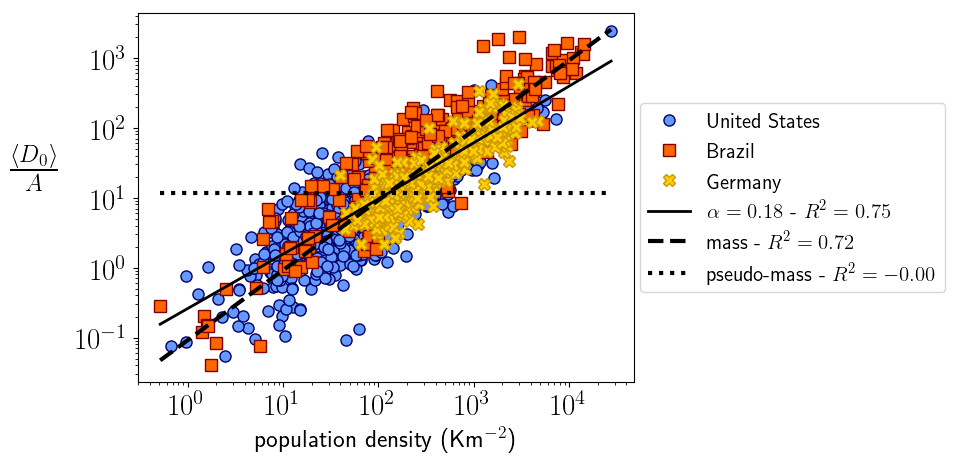}
	\caption{Comparison between $\langle D_0 \rangle / A$ and the population density for different cities around the world.}
	\label{fig:world_rho}
\end{figure}

In Figs.~\ref{fig:US_rho},~\ref{fig:BR_rho} and~\ref{fig:GE_rho} we show, respectively, the comparison between $\langle D_0 \rangle / A$ and the population density for different counties of United States, cities of Brazil, and cities of Germany. We can note that the model (in Eq~\ref{eq:model_den}) provides a good fit for United States, better than the mass action hypothesis. However, this not happens in Brazil and Germany, where the model (in Eq~\ref{eq:model_den}) have almost the same predictability that this hypothesis. The pseudo-mass-action cannot explain these results. This is also true in  a more general scope, as shown in Fig.~\ref{fig:world_rho}.

This result can indicate that the contact rate is, in fact, related with population size and not with the population density. The case of United states can be explained by the linear scaling between their county's population size and population density, as show before in Fig.~\ref{fig:pop_den_US}.

\subsection{Population size driven contact rate}

If the mass action (constant contact rate) is valid, we expect that $\langle D_0 \rangle$ scales linearly with $N$. In the another extreme, if the pseudo-mass action ($C \propto N$) is valid, we expect that $\langle D_0 \rangle$ is a constant.

\begin{figure}[!htb]
	\centering	
	\includegraphics[height=7cm]{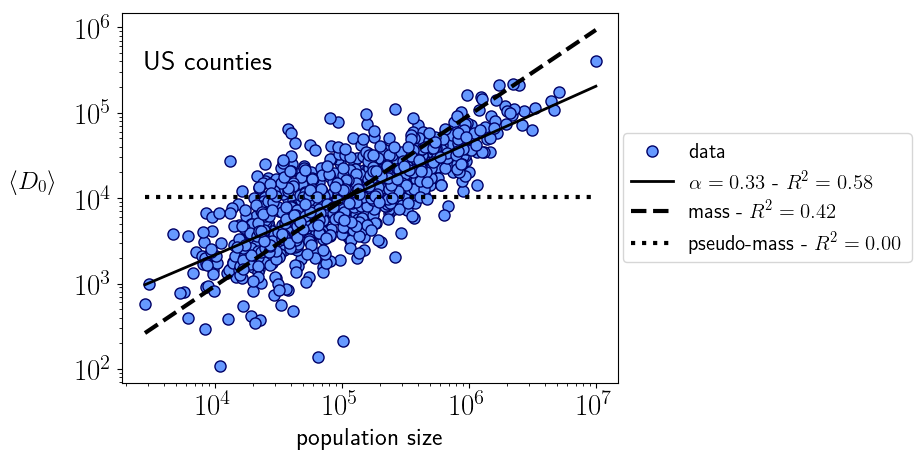}
	\caption{Comparison between $\langle D_0 \rangle$ and the population size for different counties of United States.}
	\label{fig:US_N}
\end{figure}

\begin{figure}[!htb]
	\centering	
	\includegraphics[height=7cm]{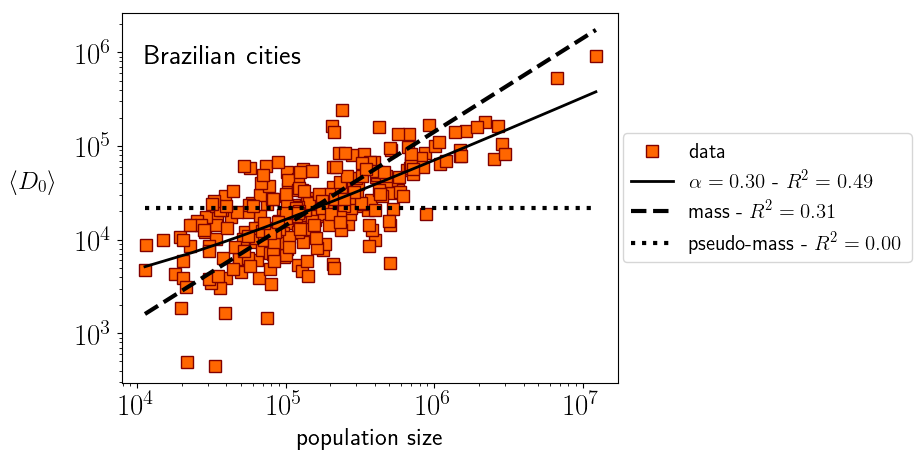}
	\caption{Comparison between $\langle D_0 \rangle$ and the population size for different cities of Brazil.}
	\label{fig:BR_N}
\end{figure}

\begin{figure}[!htb]
	\centering	
	\includegraphics[height=7cm]{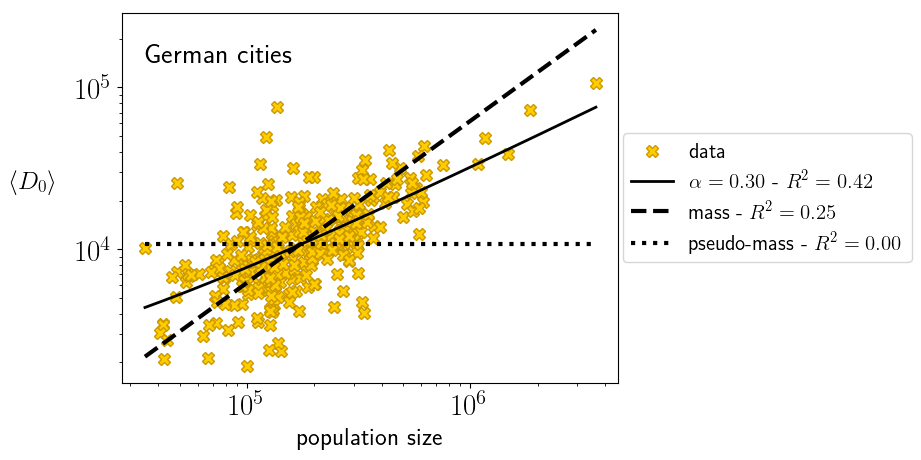}
	\caption{Comparison between $\langle D_0 \rangle$ and the population size for different cities of Germany.}
	\label{fig:GE_N}
\end{figure}

\begin{figure}[!htb]
	\centering	
	\includegraphics[height=7cm]{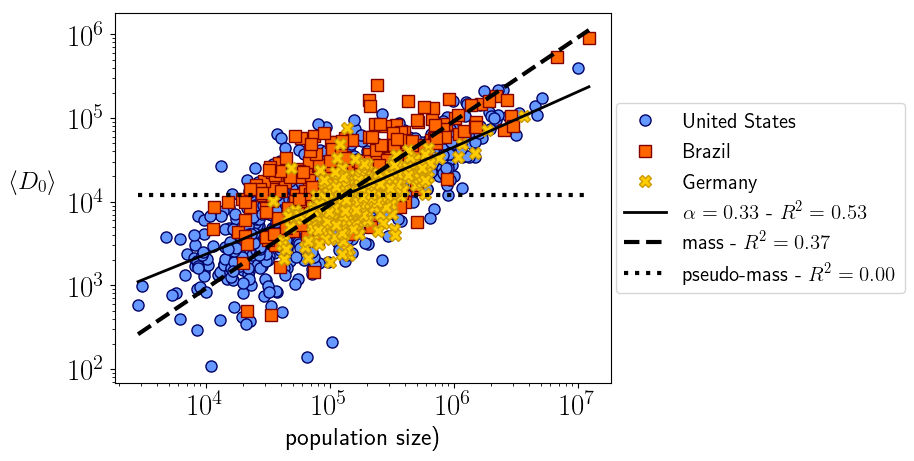}
	\caption{Comparison between $\langle D_0 \rangle$ and the population size for different cities around the world.}
	\label{fig:world_N}
\end{figure}

In Figs.~\ref{fig:US_N},~\ref{fig:BR_N} and~\ref{fig:GE_N} we show, respectively, the comparison between $\langle D_0 \rangle$ and the population size for different counties of United States, cities of Brazil, and cities of Germany. Now, we can note that the model (in Eq~\ref{eq:model_pop}) provides a good fit for the three countries and are better than the mass action hypothesis. The pseudo-mass action cannot explain these results. This can be viewed in a more general scope in Fig.~\ref{fig:world_N}, where we found universally that $\alpha \approx 1/3$. 

\section{Scaling and Scale}

\begin{figure}[!htb]
	\centering	
	\includegraphics[height=7cm]{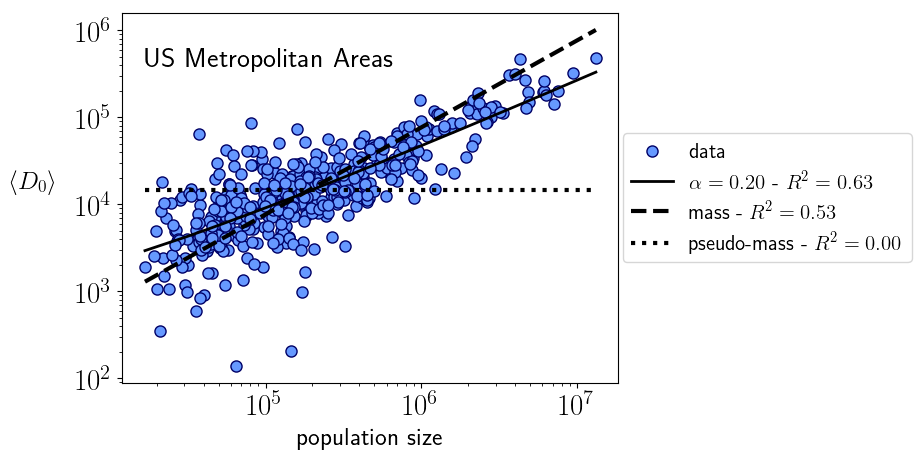}
	\caption{Comparison between $\langle D_0 \rangle$ and the population size for different Metropolitan Areas of United States.}
    \label{fig:US_N_msa}    
\end{figure}

\begin{figure}[!htb]
	\centering	
	\includegraphics[height=7cm]{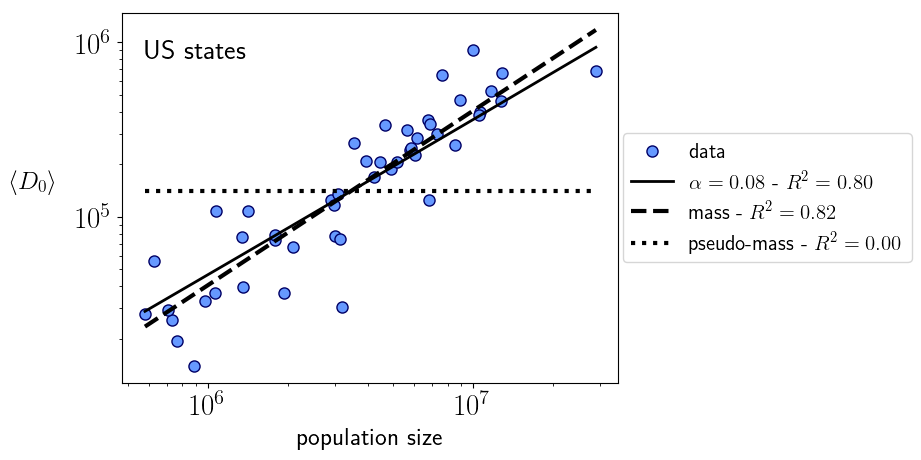}
	\caption{Comparison between $\langle D_0 \rangle$ and the population size for different States of United States.}
    \label{fig:US_N_states}
\end{figure}

Our main hypothesis is that closer people interact more frequently. So, we expect that if we increase the geographical scale (counties $\rightarrow$ metropolitan areas $\rightarrow$ states), we reduce the dependence between contact rate and the population size. In Fig.~\ref{fig:US_N_msa} and \ref{fig:US_N_states}, we respectively show the relation between $\langle D_0 \rangle$ and $N$ for Metropolitan Areas
~\footnote{We aggregate the county-level to the Metropolitan Statistical Areas using delineation files of US Office of Budget and Management
~\cite{eua_msa}.} and States of United States. As expected, the scaling dependence is higher for geographical scales with more granularity.

The scaling $\alpha \approx 0.2$ for Metropolitan Areas is very close to the power-law scaling found in a previous work~\cite{stier2020covid}. To do so, using the approximation $S \approx N$ for short-times, they measure the growth rate of confirmed cases of COVID-19 by an exponential fit between March 16th and March 25th. Here we follow the approach described in the methodology section, since it not involves approximations, uses all available data (both confirmed cases and deaths) and allows the weekly variation of $\beta$, since now we have access of a longer period.

\section{Conclusions}

The epidemic dynamics are traditionally explained by two hypothesis: the mass action and the pseudo-mass action. Here we shown empirically that neither is good to describe the data. Also, we develop a theory to explain the found relation.

Our analysis and results give support to the validity of the population size driven contact rate for the COVID-19 pandemic. This result can also explain the super-linear scaling of criminality in Brazil~\cite{gomez2012statistics}, Japan~\cite{bettencourt2013origins} and United States~\cite{oliveira2017scaling}. Such is the the downside of leaving in large urban centers.

From our analysis, it is clear that the scaling is valid at the municipal, county, or city level. If we make it broader at regions, province or state level, it is washed out by the different scales averaged over such large regions. 

This conclusion can provide useful insight regarding the urgent problem that cities, and the world in general, are facing.
As others authors~\cite{stier2020covid} already pointed out, larger cities require more strict social distancing policies. On the other side, smaller cities may relax controls before larger cities.

\bibliographystyle{unsrt}  

\bibliography{urban-covid19}

\begin{thebibliography}{10}

\bibitem{Weber2020}
Albertine Weber, Flavio Iannelli, and Sebastian Gonçalves.
\newblock Trend analysis of the covid-19 pandemic in china and the rest of the
  world.
\newblock {\em medRxiv} 2020.03.19.20037192, 2020.

\bibitem{keeling2011modeling}
Matt~J Keeling and Pejman Rohani.
\newblock {\em Modeling infectious diseases in humans and animals}.
\newblock Princeton University Press, 2011.

\bibitem{anderson-may1991}
R.M. Anderson and R.M. May.
\newblock {\em Infectious Diseases of Humans}.
\newblock Oxford University Press, 1991.

\bibitem{hu2013scaling}
Hao Hu, Karima Nigmatulina, and Philip Eckhoff.
\newblock The scaling of contact rates with population density for the
  infectious disease models.
\newblock {\em Mathematical biosciences}, 244(2):125--134, 2013.

\bibitem{bettencourt2013origins}
Lu{\'\i}s~MA Bettencourt.
\newblock The origins of scaling in cities.
\newblock {\em science}, 340(6139):1438--1441, 2013.

\bibitem{pentland2020}
C.~Krumme M.~Cebrian W.~Pan, G.~Ghoshal and A.~Pentland.
\newblock Urban characteristics attributable to density-driven tie formation.
\newblock {\em Nature Communications}, 4:1961, 2013.

\bibitem{bjornstad2002dynamics}
Ottar~N Bj{\o}rnstad, B{\"a}rbel~F Finkenst{\"a}dt, and Bryan~T Grenfell.
\newblock Dynamics of measles epidemics: estimating scaling of transmission
  rates using a time series sir model.
\newblock {\em Ecological monographs}, 72(2):169--184, 2002.

\bibitem{noulas2012tale}
Anastasios Noulas, Salvatore Scellato, Renaud Lambiotte, Massimiliano Pontil,
  and Cecilia Mascolo.
\newblock A tale of many cities: universal patterns in human urban mobility.
\newblock {\em PloS one}, 7(5), 2012.

\bibitem{CSSEGI}
Cssegi sand data / covid-19.
\newblock \url{https://github.com/CSSEGISandData/COVID-19}.
\newblock Accessed: 2020-05-15.

\bibitem{brasilio}
Brasil io / covid-19.
\newblock \url{https://brasil.io/covid19/}.
\newblock Accessed: 2020-05-15.

\bibitem{jgehrcke}
Covid-19 germany gae.
\newblock \url{https://github.com/jgehrcke/covid-19-germany-gae/}.
\newblock Accessed: 2020-05-15.

\bibitem{eua_pop}
County population totals: 2010-2019.
\newblock
  \url{https://www.census.gov/data/datasets/time-series/demo/popest/2010s-counties-total.html}.
\newblock Accessed: 2020-05-15.

\bibitem{eua_area}
2010 census urban and rural classification and urban area criteria.
\newblock
  \url{https://www.census.gov/programs-surveys/geography/guidance/geo-areas/urban-rural/2010-urban-rural.html}.
\newblock Accessed: 2020-05-15.

\bibitem{brasil_pop}
Estimativas da população.
\newblock
  \url{https://www.ibge.gov.br/estatisticas/sociais/populacao/9103-estimativas-de-populacao.html}.
\newblock Accessed: 2020-05-15.

\bibitem{brasil_area}
Área territorial 2018.
\newblock
  \url{https://www.ibge.gov.br/geociencias/organizacao-do-territorio/estrutura-territorial/15761-areas-dos-municipios.html}.
\newblock Accessed: 2020-05-15.

\bibitem{pause-ny}
No. 202: Declaring a disaster emergency in the state of new york.
\newblock
  \url{https://www.governor.ny.gov/news/no-202-declaring-disaster-emergency-state-new-york}.
\newblock Accessed: 2020-05-15.

\bibitem{eua_msa}
Metropolitan and micropolitan.
\newblock
  \url{https://www.census.gov/programs-surveys/metro-micro/about/delineation-files.html}.
\newblock Accessed: 2020-05-15.

\bibitem{stier2020covid}
Andrew Stier, Marc Berman, and Luis Bettencourt.
\newblock Covid-19 attack rate increases with city size.
\newblock {\em Mansueto Institute for Urban Innovation Research Paper
  Forthcoming}, 2020.

\bibitem{gomez2012statistics}
Andres Gomez-Lievano, HyeJin Youn, and Luis~MA Bettencourt.
\newblock The statistics of urban scaling and their connection to zipf’s law.
\newblock {\em PloS one}, 7(7), 2012.

\bibitem{oliveira2017scaling}
Marcos Oliveira, Carmelo Bastos-Filho, and Ronaldo Menezes.
\newblock The scaling of crime concentration in cities.
\newblock {\em PloS one}, 12(8), 2017.

\end{thebibliography}

\end{document}